\documentclass{ws-procs975x65}

\newcommand{\lanln}[1]{$\langle$\texttt{arXiv:#1}$\rangle$}

\begin{document}

\title{BLACK HOLE ENTROPY AND THE RENORMALIZATION GROUP}

\author{ALEJANDRO SATZ$^*$ and TED JACOBSON$^\dagger$}

\address{Physics Department, University of Maryland,\\
College Park, Maryland 20742, USA\\
$^*$E-mail: alesatz@umd.edu\\
$^\dagger$E-mail: jacobson@umd.edu}

\begin{abstract}
Four decades after its first postulation by Bekenstein, black hole
entropy remains mysterious. It has long been suggested that the
entanglement entropy of quantum fields on the black hole gravitational
background should represent at least an important contribution to the
total Bekenstein-Hawking entropy, and that the divergences in the
entanglement entropy should be absorbed in the renormalization of the
gravitational couplings. In this talk, we describe how an improved
understanding of black hole entropy is obtained by combining these
notions with the renormalization group. By introducing an RG flow
scale, we investigate whether the total entropy of the black hole can
be partitioned in a ``gravitational'' part related to the flowing
gravitational action, and a ``quantum'' part related to the unintegrated
degrees of freedom. We describe the realization of this idea for free
fields, and the complications and qualifications arising for
interacting fields.\end{abstract}


\bodymatter

\section{Entanglement entropy and black hole entropy}\label{aba:sec1}

The global Hartle-Hawking vacuum state on the Schwarzschild black hole, restricted to the exterior of the horizon, is a themal state at the Hawking temperature $\beta_H$\cite{israel}. Hence the horizon entanglement entropy matches the thermal entropy of the fields at the exterior \cite{kabat
%solod1
}. It can be computed from a Euclidean partition function:
\begin{equation}
S_{\mathrm{ent}}=-\left(\beta\partial_\beta-1\right)\left.\ln Z(\beta)\right|_{\beta=\beta_H}\,,
\end{equation}
where the $\beta$-derivative introduces a conical singularity at the horizon\cite{suss}. However, the same result can be achieved with an ``on-shell'' computation, starting from a path integral over metrics and matter fields with suitable boundary conditions, and approximating the gravitational path integral by the saddle-point evaluation at the solution $\bar{g}(\beta)$ of the effective equations of motion:
\begin{equation}
Z(\beta)=\int\mathcal{D}g\int\mathcal{D}\varphi\,\mathrm{e}^{-S_b[g]-S[g,\varphi]}=\int\mathcal{D}g\,\mathrm{e}^{-\Gamma_0[g]}\approx\mathrm{e}^{-\Gamma_0[\bar{g}(\beta)]}\,.
\end{equation}
Then $(\beta\partial_\beta-1)\Gamma_0[\bar{g}(\beta)]$ gives the total black hole entropy, and the term of it representing the matter contribution equals the entanglement entropy because the ``on-shell'' and the ``off-shell'' $\beta$-derivatives coincide\cite{solod3}. In a one-loop approximation, the matter contribution to effective action $\Gamma_0$ can be computed with a heat-kernel expansion using a UV regulator; its divergences can then be absorbed into a renormalization of the bare gravitational couplings in $S_b$ \cite{suss2, wilzec}. The purpose of this contribution is to exhibit this renormalization property of the entropy using the Wilsonian renormalization group \cite{wilson} to avoid working with divergences and infinite renormalizations. Many of the results are also derived and discussed at length in a companion paper\cite{bheerg}.

\section{RG and black hole entropy: free fields}

Let us consider first the minimally coupled scalar field. We define the gravitational effective action at a Wilsonian scale $k$ by:
\begin{equation}
\mathrm{e}^{-\Gamma_k[g]}=\mathrm{e}^{-S_b[g]}\int\mathcal{D}\varphi\,\mathrm{e}^{-\frac{1}{2}\int\varphi(-\nabla^2+\mathcal{R}_k(\nabla^2))\varphi}\,,
\end{equation}
where $\mathcal{R}_k$ is an IR cutoff function, suppressing the momenta  $p <k$ and vanishing for  $p>k$. The full effective action is then given by
\begin{equation}\label{gammas}
\Gamma_0[g]=\Gamma_k[g]-\ln\int \mathcal{D}\varphi\,\mathrm{e}^{-\frac{1}{2}\int\varphi\left(\frac{-\nabla^2}{-\nabla^2+\mathcal{R}_k}\right)\varphi}=\Gamma_k[g]+\frac{1}{2}\mathrm{Tr}\ln\left[\frac{-\nabla^2}{-\nabla^2+\mathcal{R}_k}\right]\,.
\end{equation}
The second term of the right hand side contains an inbuilt UV cutoff at scale $k$, so it represents the contribution to $\Gamma_0$ of the modes below $k$. The black hole entropy is computed using a heat kernel expansion to evaluate these quantities, evaluating on the Euclidean Schwarzschild metric $\bar{g}(\beta)$, and applying the $(\beta\partial_\beta-1)$ operator. The result, to the lowest order in curvature, is:
\begin{equation}
S_{BH}=\frac{A_\Sigma}{4G_0}=\frac{A_\Sigma}{4G_k}+\frac{A_\Sigma k^2}{48\pi}\,.
\end{equation}
Thus the total entropy gets partitioned into an ``effective gravitational'' contribution expressed in terms of the running Newton constant, and a term matching the expected expression for the entanglement entropy of the degrees of freedom below scale $k$. (However, the partition of degrees of freedom at scale $k$ is defined here in the Euclidean path integral and lacks a straightforward Lorentzian interpretation.)

An analogous computation for the nonminimally coupled scalar field gives:
\begin{equation}
S_{BH}=\frac{A_\Sigma}{4G_0}=\frac{A_\Sigma}{4G_k}+\frac{A_\Sigma k^2}{8\pi}\left(\frac{1}{6}-\xi\right)\,.
\end{equation}
Here the $\xi$-term is not interpretable as entanglement entropy but as Wald entropy \cite{wald, frolov}, being equal to $2\pi\xi\int_\Sigma\langle\varphi^2\rangle$ computed with a $UV$ cutoff at scale $k$.

\section{Interacting fields}
In the case of interacting fields, we need to keep track of the nontrivial running of the matter action caused by integrating the upper modes. We do this through the Polchinski RG flow equation\cite{polch}. The partition function can be expressed as a path integral over the modes below scale $k$ (with normalization factor $N_k$):
\begin{equation}
Z=N_k\int\mathcal{D}\varphi\,\mathrm{e}^{-\frac{1}{2}\int\varphi\frac{1}{P_k}\varphi-S_k[g,\varphi]}\,,
\end{equation}
using the Wilsonian effective action $S_k$, defined by the path integral over modes above scale $k$, and the propagator $P_k$, which is UV-cutoffed at scale $k$. We define the  gravitational effective action, $\Gamma_k=S_k[g,\varphi=0]$ and the non-kinetic terms of the effective action for $\varphi$, $\tilde{S}_k[g,\varphi]=S_k[g,\varphi]-\Gamma_k[g]$. The entropy is expressed as
\begin{equation}\label{entropiesint}
 S_{BH}=(\beta\partial_\beta-1) \Gamma_k[\bar{g}]\,-(\beta\partial_\beta-1) \ln \left[ N_k \int\mathcal{D}\varphi\,\mathrm{e}^{-\frac{1}{2}\int\varphi P^{-1}_k \varphi\,-\,\tilde{S}_k[\bar{g},\varphi]}\right]\,,
\end{equation}
where the first term represents the effective gravitational entropy at scale $k$, and the remaining term the contribution of the matter fields below $k$. The flow of each of these terms can be computed solving the Polchinski flow equation,
\begin{equation}
\dot{S}_k[g,\varphi]=\frac{1}{2}\left\{\frac{\delta S_k}{\delta\varphi}\cdot \dot{P}_k\cdot\frac{\delta S_k}{\delta \varphi}-\mathrm{Tr}\left[\dot{P}_k\cdot\left(\frac{\delta^2S_k}{\delta\varphi\delta\varphi}- \frac{1}{P_k+(-\nabla^2)^{-1}}\right)\right]\right\}\,,
\end{equation}
(where the overdot is a $k$-derivative).  However, this contribution of the second term in (\ref{entropiesint}) cannot be identified in a naive way with ``the entropy of the lower modes'' as a subsystem, because in an interacting theory the modes below $k$ are entangled with the modes above $k$ \cite{momentum}. This point is discussed at length elsewhere\cite{bheerg}.

\section{Summary and outlook}

We have studied the black hole entropy $S_{BH}$ in semiclassical gravity, defining it through a canonical partition function. Introducing a remormalization group scale $k$, the full effective action $\Gamma_0$ from which $S_{BH}$ is computed gets separated in two contributions: the gravitational effective action at scale $k$, from which an effective gravitational entropy is expressed in terms of the running Newton constant, and the path integral over field modes below scale $k$. In the minimally coupled free field case, the latter contribution can be identified with the entanglement entropy of the (Euclidean) modes below $k$; the interpretation is less clear for interacting fields. Further research is needed to clarify better the interacting case and to make contact with the computations of the gravitational effective action done using the Exact Renormalization Group, which have been also related to black hole entropy\cite{
%litimfalls,
reuterbecker, litim}.

\bibliographystyle{ws-procs975x65}
\bibliography{ws-pro-sample}

\end{document}